\documentclass{acm_proc_article-sp}

\usepackage{tikz}
\usetikzlibrary{arrows,positioning,fit,calc,decorations.markings}
\usepackage{pgfplots}
\pgfplotsset{compat=newest} 
\usepgfplotslibrary{units} 
\usepackage{multirow}
\usepackage{url}
\usepackage{listings}
\usepackage{placeins}
\newcommand{\code}[1]{\texttt{#1}}
\newcommand{\elq}{ELQ}
\newcommand{\gls}[1]{#1}

\begin{document}

\title{Modular Responsive Web Design using Element Queries}
\subtitle{}
%
%
%
%
%

\numberofauthors{3} 
%
\author{
%
%
\alignauthor
  Lucas Wiener\\
  \affaddr{EVRY}\\
  \affaddr{Stockholm, Sweden}\\
  \email{lucas.wiener@evry.com}
\alignauthor
  Tomas Ekholm\\
  \affaddr{KTH Royal Institute of Technology}\\
  \affaddr{Stockholm, Sweden}\\
  \email{tomase@kth.se}
\alignauthor
  Philipp Haller\\
  \affaddr{KTH Royal Institute of Technology}\\
  \affaddr{Stockholm, Sweden}\\
  \email{phaller@kth.se}
}

\date{10 October 2015}

\maketitle
\begin{abstract}
  Responsive Web Design (RWD) enables web applications to adapt to the characteristics of different devices such as screen size which is important for mobile browsing.
  Today, the only W3C standard to support this adaptability is CSS media queries.
  However, using media queries it is impossible to create applications in a modular way, because responsive elements then always depend on the global context.
  Hence, responsive elements can only be reused if the global context is exactly the same, severely limiting their reusability.
  This makes it extremely challenging to develop large responsive applications, because the lack of true modularity makes certain requirement changes either impossible or expensive to realize.

  In this paper we extend RWD to also include responsive modules, i.e., modules that adapt their design based on their local context independently of the global context.
  We present the \elq{} project which implements our approach.
  \elq{} is a novel implementation of so-called \emph{element queries} which generalize media queries.
  Importantly, our design conforms to existing web specifications, enabling adoption on a large scale.
  \elq{} is designed to be heavily extensible using plugins.
  Experimental results show speed-ups of the core algorithms of up to 37x compared to previous approaches.

\end{abstract}





\newpage
\section{Introduction}

  Responsive Web Design (RWD) is an approach to make an application respond to the viewport size and device characteristics.
  This is currently achieved by using CSS media queries that are designed to conditionally design content by the media, such as using serif fonts when printed and sans-serif when viewed on a screen \cite{w3c_css_mq}.

  In order to reduce complexity and enable reusability, applications are typically composed of modules, i.e., interchangeable and independent parts that have a single and well-defined responsibility \cite{parnas1972criteria}.
  In order for a module to be reusable it must not assume in which context it is being used.

  In this paper we focus on the presentation layer of web applications.
  As it stands, using media queries to make the presentation layer responsive precludes modularity.
  The problem is that there is no way to make a module responsive without making it context-aware, due to the fact that media queries can only target the viewport; this means that responsive elements can only respond to changes of the (global) viewport.
  Thus, a responsive module using media queries is layout dependent and has both reduced functionality and limited reusability~\cite{elq-thesis}.
  As a result, media queries can only be used for RWD of non-modular static applications.
  In a world where no better solution than media queries exists for RWD, changing the layout of a responsive application becomes a cumbersome task.

  \subsection{The Problem Exemplified}
    
    Imagine an application that displays the current weather of various cities as widgets, by using a weather widget module.
    The module should be responsive so that more information, such as a temperature graph over time, is displayed when the widget is big.
    When the widget is small it should only display the current temperature.
    Users should also be able to add, remove and resize widgets.

    Such an application cannot be built using media queries, since the widgets can have varying sizes independent of the viewport (e.g., the width of one widget is 30\% while another is 40\%).
    To overcome this problem we must change the application, so that widgets always have the same sizes.
    This implies that the size of the module and the media query breakpoints are coupled/intertwined, i.e. they are proportional to each other.
    The problem now is that we have removed the reusability of the weather module, since it requires the specific width that is correctly proportional to the media query breakpoints. 

    Imagine a company working on a big application that uses media queries for responsiveness (i.e., each responsive module assumes to have a specific percentage of the viewport size).
    The ability to change is desired by both developers and stakeholders, but is limited by this responsive approach.
    The requirement of changing a menu from being a horizontal menu at the top to being a vertical menu on the side implies that all responsive modules break, since the assumed proportionality of each module is changed.
    Even worse, if the menu is also supposed to hide on user input, the responsiveness of the module breaks, since the layout changes dynamically.
    The latter requirement is impossible to satisfy in a modular way without element queries.

    Additionally, it is popular to define breakpoints relative to the font size so that conditional designs respect the size of the content~\cite{mq-em}.
    Media queries can only target the font size of the document root, limiting their functionality drastically.
    With element queries breakpoints may be defined relative to the font size of the targeted element.

    As we can see, even with the exemplified limited requirements there are still significant restrictions when using media queries for responsive modules.
    
    \subsection{Requirements}\label{sec:reqs}

      The desired behavior of a responsive module is having its inner design respond to the size of \emph{its container} instead of the viewport.
      Only then is a responsive module independent of its layout context.
      Realizing responsive modules requires CSS rules that are conditional upon \emph{elements}, instead of the global viewport.
      We have identified the following requirements of a solution:

      \begin{itemize}
        \item 
          It must provide the possibility for an element to automatically respond to changes of its parent's properties.
        \item
          It must conform to the syntax of HTML, CSS, and JavaScript to retain the compatibility of tools, libraries and existing projects.
        \item
          It must have adequate performance for large applications that make heavy use of responsive modules.
        \item
          It must enable developers to write encapsulated style rules, so that responsive modules may be arbitrarily composed without any conflicting style rules.
      \end{itemize}

    \subsection{Approach}
      In this paper we extend the concept of RWD to also include responsive modules.
      The W3C has discussed such a feature under the name of \emph{element queries} given its analogy to media queries~\cite{w3c_eq_mail}.
      This paper presents a novel implementation of element queries in JavaScript named \elq{} that enables new possibilities of RWD.
      Our approach satisfies all requirements given in Section~\ref{sec:reqs}.
      We have released ELQ as an open-source library under the MIT license.\footnote{\url{https://github.com/elqteam/elq}}
      The implementation supports all major browsers, including Internet Explorer version 8, Chrome version 42 (the last version compatible with Android version 4), Safari version 5, and Opera version 12.

      One could argue that a solution does not need to be executed on the client side, but instead generate media queries on the server side for all modules with respect to the current application layout.
      However, this approach is insufficient, since it limits modules to applications with static layouts~\cite{elq-thesis}. Also, the generated media queries would not be able to respond to the user changing properties of elements such as layout and font size.

    \subsection{Contributions}
      This paper makes the following contributions:
      \begin{itemize}
        \item A new design for element queries that enables responsive modules while conforming to the syntax of HTML, CSS, and JavaScript.
        \item
          Our approach is the first to enable nested elements that are responsive in a modular way, i.e., modules fully encapsulate any styling required for RWD.
          As a side effect, responsive modules may also be arbitrarily styled with CSS independent of their context.
        \item
          An extensible architecture that enables plugins to significantly extend the behavior of ELQ, our library implementation.
          This makes it possible to create plugins in order to enable new features and to ease integration of \elq{} into existing projects.
        \item
          A new implementation that offers substantially higher performance than previous approaches.
          The implementation batch-processes DOM operations in order to avoid layout thrashing (i.e., forcing the layout engine to perform multiple independent layouts).
        \item
          A run-time cycle detection system that detects and breaks cycles stemming from cyclic rules due to unrestricted usage of element queries~\cite{elq-thesis}.
      \end{itemize}

    The rest of the paper is organized as follows.
    Section~\ref{sec:elq} introduces \elq{} and its API from a user's perspective.
    In Section~\ref{sec:plugins} we introduce \elq{}'s plugin architecture.
    Section~\ref{sec:imp} provides an overview of the main components of \elq{}'s implementation.
    In Section~\ref{sec:eval} we evaluate the performance of \elq{} and report on case studies.
    Section~\ref{sec:discussion} discusses limitations of \elq{} and related libraries, as well as the current state of standardization of element queries.
    Section~\ref{sec:related} relates \elq{} to prior work, and Section~\ref{sec:conclusion} concludes.

\section{Overview of \elq{}}\label{sec:elq}
  Media queries and element queries are similar in the sense that they both enable developers to define conditional designs that are applied by specified criteria.
  The main difference is the type of criteria that can be used.
  With media queries critera of the device, document, and media are used, while element criteria are used with element queries.
  It can somewhat simplified be described as that media queries target the document root and up such as viewport, browser, device, and input mechanisms.
  Element queries target the document root and down, i.e., elements of the document.

  \elq{} is designed to be plugin-based for increased flexibility and extensibility.
  By providing a good library foundation and plugins it is up to developers to choose the right plugins for each project.
  In addition, by letting the plugins satisfy the requirements it is easy to extend the library with new plugins when new requirements arise.

  An \emph{element breakpoint} is defined as a point of an element property range which can be used to define conditional behavior, similar to breakpoints of media queries.
  For example, an element breakpoint of 500 pixels in width enables conditional styling depending on if the element is narrower or wider than 500 pixels.
  An element may have multiple breakpoints.
  An \emph{element breakpoint state} is defined as the state of the element breakpoint relative to the current element property value.
  For example, if an element that is 300 pixels wide has two width breakpoints of 200 and 400 pixels the element breakpoint states are ``wider than 200 pixels'' and ``narrower than 400 pixels''.

  When the breakpoint states of an element changes, \elq{} performs cycle detection in order to detect and handle possible style cycles.
  If a cycle is detected, the new element breakpoint states are not applied in order to avoid an infinite loop of layouts.
  The cycle detection system is implemented as an conservative algorithm, and may in some cases detect false positives.

  \subsection{The API}\label{sec:elq-api}
    In this section, we use the \code{elq-breakpoints} API (that is bundled with \elq{} as default) that let use define element breakpoints.
    The main idea is to define element breakpoints of interest so that children can be conditionally styled in CSS by targeting the different element breakpoint states.
    As CSS3 does not support custom at-rules/selectors \cite{w3c_css_selectors}, responsive elements are annotated in HTML by element attributes.
    \elq{} then observes the annotated elements in order to automatically update breakpoint state classes.
    Although not written in the examples, the API also supports attributes defined with the \code{data-} prefix to conform to the HTML standard \cite{html-spec}.

    The following example shows the HTML of an element that has two annotated width breakpoints at 300 and 500 pixels:

    \begin{lstlisting}[gobble=6,caption={},captionpos=b,label={}]]
      <div class="foo" elq elq-breakpoints
        elq-breakpoints-widths="300 500">
        
        <p>When in doubt, mumble.</p>
      </div>
    \end{lstlisting}

    When \elq{} has processed the element it will have two classes that reflect each breakpoint state.
    For instance, if the element is 400 pixels wide, the element has the two classes \code{elq-min-width-300px} and \code{elq-max-width-500px}.
    Similarly, if the element is 200 pixels wide the classes are instead \code{elq-max-width-300px} and \code{elq-max-width-500px}.
    So for each breakpoint only the min/max part changes.
    It may seem alien that the classes describe that the width of the element is both maximum 300 and 500 pixels.
    This is because we have taken a user-centric approach so that the CSS usage of the classes is similar to the API of media queries.
    However, developers are free to change this API by creating plugins.

    Now that we have defined the breakpoints of the element, we can conditionally style it in CSS by using the classes as shown in listing~\ref{code:elq-breakpoints-example-css}.

    \begin{lstlisting}[gobble=6,caption={Example usage of the breakpoint state classes in CSS.},captionpos=b,label={code:elq-breakpoints-example-css}]]
      .foo.elq-max-width-300px {
        background-color: blue;
      }

      .foo.elq-min-width-300px.elq-max-width-500px {
        background-color: green;
      }

      .foo.elq-min-width-500px p {
        color: white;
      }
    \end{lstlisting}

    In order for the conditional styles to be applied, the elements that have breakpoints must be activated by the \elq{} JavaScript runtime.
    \elq{} can either be required as a module (by the CommonJS or AMD syntax), or it can be included in a HTML \code{script} tag which then will expose a global constructor \code{Elq}.
    The following is an example of how to create an \elq{} instance and activating elements:

    \begin{lstlisting}[gobble=6,caption={}]]

      // Create an instance
      var elq = Elq();

      // Plugins may be registered
      elq.use(myPlugin);
      elq.use(myOtherPlugin);

      // Activate elements.
      var elements = document.querySelectorAll("[elq]");
      elq.activate(elements);

    \end{lstlisting}

    In this example we create an \elq{} instance and register two plugins with it.
    Then we query the document for all elements with an \code{elq} attribute (as annotated in the previous example) and then pass them as an argument to the activation method of \elq{}.
    It should be noted that it is up to developers how to activate elements; annotating elements with \code{elq} is used for simplicity in the example.
    The only requirement is that conditionally-styled elements are processed by the \code{activate} method at some point.
    This can, for example, also be achieved with a plugin that listens to DOM mutations to perform the activation automatically, or a plugin that parses CSS and activates all elements that have conditional styles defined.

    \subsubsection{Nested modules}
      The \code{elq-breakpoints} API is sufficient for applications that do not need nested breakpoint elements, and similar features are provided by related libraries such as \cite{eq_imp_eqjs,eq_imp_responsive-elements-2}.
      However, using such an API in responsive modules still limits composability, since modules then may not exist in an outer responsive context.

      The reason this API is not sufficient for nested modules is that there is no way to limit the CSS matching search of the selectors.
      The last style rule of the example given in listing~\ref{code:elq-breakpoints-example-css} specifies that all paragraph elements should have white text if \emph{any} ancestor breakpoints element is wider than 500 pixels.
      Since the ancestor selector may match elements outside of the module, such selectors are dangerous to use in the context of responsive modules.
      The problem may be somewhat reduced by more specific selectors and such, but it cannot be fully solved for arbitrary styling \cite{elq-thesis}.

      To solve this problem, we provide a plugin that let us define elements to ``mirror'' the breakpoints classes of the nearest ancestor breakpoints element (the target of the mirror element).
      This means that the mirror element always reflects the element breakpoint states of the target.
      Then, the conditional style of the mirror element may be written as a combinatory selector that is relative to the nearest ancestor breakpoints element.
      The following is an example usage of the mirror plugin to enable nested modules:

      \begin{lstlisting}[gobble=8,caption={},captionpos=b,label={}]]
        <div class="foo" elq elq-breakpoints
          elq-breakpoints-widths="300 500">
          
          <div class="foo" elq elq-breakpoints
            elq-breakpoints-widths="300 500">

            <p elq elq-mirror>...</p>
          </div>

          <p elq elq-mirror>...</p>
        </div>
      \end{lstlisting}

      In this example, the paragraph elements always have the same element breakpoint classes as the parent \code{elq-breakpoints} elements.
      This enables us to write CSS that does not traverse the ancestor tree:

      \begin{lstlisting}[gobble=8,caption={},captionpos=b,label={code:elq-mirror-example-css}]]
        .foo {
          /* So that the nested modules
             have different size */
          width: 50%;
        }

        .foo p.elq-min-width-500px {
          color: white;
        }
      \end{lstlisting}

      In the examples we have given so far we have annotated element breakpoints manually; however, this does not properly show the power and flexibility of \elq{}'s API. Therefore, the next section presents an API that combines JavaScript and generated CSS in order to create a flexible grid API.

    \subsubsection{A grid API}
      In this section we present a plugin that defines an API that enables developers to use responsive grids consisting of twelve columns and utility classes very similar to the ones defined by the CSS Bootstrap framework \cite{bootstrap}.
      The goal of the API is to provide an abstraction of element queries, so that developers may focus on responsivity using classes instead of the syntax presented in previous sections.

      The following is an example grid:

      

      
      \begin{lstlisting}[gobble=8,caption={},captionpos=b,label={}]]
        <div class="container">
          <div class="row">
            <div class="col-500-4 col-700-6">
              ...
            </div>
            <div class="col-500-4" col-700-6>
              ...
            </div>
            <div class="col-500-4 hidden-700-up">
              ...
            </div>
          </div>
        </div>
      \end{lstlisting}

      The example grid is defined to be single columned when the width of the grid is below 500 pixels, triple columned when the width is between 500 and 700 pixels, and double columned for when the width is above 700 pixels.
      The last column is hidden when the width is above 700 pixels.

      The column classes define the behaviour of the grid, and have the syntax \code{col-[breakpoint]-[size]}.
      The \code{[breakpoint]} part of a column class is relative to the parent \code{row} and can be any positive number including an optional unit.
      Currently, the supported units are \code{px, em, rem}.
      If the unit is omitted, \code{px} is assumed.
      Grids may also be nested.

      The plugin traverses the grid structure to initialize all columns and possible nested grids.
      It also generates and applies the CSS needed for each column breakpoint automatically to the document.
      This enables developers to easily create responsive grids in nestable modules.

\section{Extensions via plugins}\label{sec:plugins}
  For example, if annotating HTML is undesired it is possible to create a plugin that instead generates element breakpoints by parsing CSS.
  Likewise, if adding breakpoint state classes to elements is undesired it is possible to create a plugin that does something else when an element breakpoint state has changed.

  A plugin is defined by a \emph{plugin definition object} and has the structure shown in listing~\ref{code:elq-plugin-definition}.
  \begin{lstlisting}[gobble=4,caption={The structure of plugin definition objects.},captionpos=b,label={code:elq-plugin-definition}]]
    var myPluginDefinition = {
      getName: function () {
        return "my-plugin";
      },
      getVersion: function () {
        return "0.0.0";
      },
      isCompatible: function (elq) {
        return true;
      },
      make: function (elq, options) {
        return {
          // Implement plugin instance methods.
          ...
        };
      }
    };
  \end{lstlisting}

  All of the methods are invoked when registered to an \elq{} instance.
  The \code{getName} and \code{getVersion} methods tells the name and version of the plugin.
  The \code{isCompatible} tells if the plugin is compatible with the \elq{} instance that it is registered to.
  In the \code{make} method the plugin may initialize itself to the \elq{} instance and return an object that defines the plugin API accessible by \elq{} and other plugins.

  \elq{} invokes certain methods of the plugin API, if implemented, to let plugins decide the behavior of the system.
  Those methods are the following:
  \begin{itemize}
    \item \code{activate(element)}
          Called when an element is requested to be activated, in order for plugins to initialize listeners and element properties.
    \item \code{getElements(element)}
          Called in order to let plugins reveal extra elements to be activated in addition to the given element.
    \item \code{getBreakpoints(element)}
          Called to retrieve the current breakpoints of an element.
    \item \code{applyBreakpointStates(element, breakpointStates)}
          Called to apply the given element breakpoint states of an element.
  \end{itemize}

  In addition, plugins may also listen to the following \elq{} events:
  \begin{itemize}
    \item \code{resize(element)}
          Emitted when an \elq{} element has changed size.
    \item \code{breakpointStatesChanged(element, breakpointStates)}
          Emitted when an element has changed element breakpoint states (e.g., when the width of an element changed from being narrower to being wider than a breakpoint).
  \end{itemize}

  There are two main flows of the \elq{} system; activating an element and updating an element.
  When \elq{} is requested to activate an element, the following flow occurs:

  \begin{enumerate}
    \item Initialize the element by installing properties and a system that handles listeners.
    \item 
          Call the \code{getElements} method of all plugins to retrieve any additional elements to activate.
          Perform an activation flow for all additonal elements.
    \item Call the \code{activate} method of all plugins, so that plugin-specific initialization may occur.
    \item If any plugin has requested \elq{} to detect resize events of the element, install a resize detector to the element.
    \item Pass the element through the update flow.
  \end{enumerate}

  The update flow is as follows:
  \begin{enumerate}
    \item Call the \code{getBreakpoints} method of all plugins to retrieve the breakpoints of the element.
    \item Calculate the breakpoint states of the element.
    \item If any state has changed since the previous update:
    \begin{enumerate}
      \item Perform cycle detection. If a cycle is detected, then abort the flow and emit a warning.
      \item Call the \code{applyBreakpointStates} method of all plugins in order for plugins to apply the new element breakpoint states.
      \item Emit an \code{breakpointStatesChanged} event.
    \end{enumerate}
  \end{enumerate}

  Of course, there are options to disable some of the steps such as cycle detection and applying breakpoint states.
  In additon to being triggered by the activation flow and plugins, the update flow is also triggered by element resize events.

  Plugins may also use an extended API of \elq{} that offers access to subsystems such as the plugin handler, cycle detector, batch processor, etc.
  The extended API is exposed to plugins as an argument to the \code{make} method of the plugin definition object.
  In addition, plugins may set behavior properties of an element by the \code{element.elq} property.
  It is also possible for plugins to define own behavior properties for inter-plugin collaboration, or for storing plugin-specifc element state.
  Examples of behavior properties of the \elq{} core are:
  \begin{itemize}
    \item \code{resizeDetection} Indicates if resize detection should be performed.
    \item \code{cycleDetection} Indicates if cycle detection should be performed.
    \item \code{updateBreakpoints} Indicates if the element should be passed through the update flow.
    \item \code{applyBreakpointStates} Indicates that plugins may apply breakpoint states of the element (for some elements it is only necessary to emit element breakpoint state changes, without applying them to the actual element~\cite{elq-thesis}).
  \end{itemize}

  \subsection{Example Plugin Implementation}
    The \code{elq-breakpoints} API that enables developers to annotate breakpoints in HTML, as described in Section~\ref{sec:elq-api}, is implemented as two plugins.
    This shows that even the core functionality of \elq{} is implemented in terms of plugins.
    The first plugin parses the breakpoints of the element attributes.
    The second plugin applies the breakpoint states as classes.

    The following is a simplified implementation of the \code{make} method (see listing~\ref{code:elq-plugin-definition}) of the parsing plugin:

    \begin{lstlisting}[gobble=6,caption={},captionpos=b,label={}]]
      function activate(element) {
        if (!element.hasAttribute("elq-breakpoints")) {
          return;
        }

        element.elq.resizeDetection       = true;
        element.elq.updateBreakpoints     = true;
        element.elq.applyBreakpointStates = true;
        element.elq.cycleDetection        = true;
      }

      function getBreakpoints(element) {
        // Parse the "elq-breakpoints-*" attributes
        // and retrieve their breakpoints.
        return ...;
      }

      // Return the plugin API
      return {
        activate: activate,
        getBreakpoints: getBreakpoints
      };
    \end{lstlisting}

    In the \code{activate} method the plugin registers that resize detection is needed for the element and that it should be passed through the update flow.
    It also enables the application of breakpoint states and run-time cycle detection.
    Although not shown in the simplifed implementation, \code{applyBreakpointStates} and \code{cycleDetection} are in some cases disabled.

    The plugin that applies the element breakpoint states simply implements the \code{applyBreakpointStates} method to alter the \code{className} property of the element using the given element breakpoint states.

\section{Implementation}\label{sec:imp}
  \subsection{Batch Processing}\label{sec:imp_batch_processor}
    Batch processing is the foundation of the performance gains of our approach, and is therefore used by several subsystems.
    \elq{} uses a leveled batch processor, which is implemented as a stand-alone project.\footnote{\url{https://github.com/wnr/batch-processor}}
    It serves two purposes: to process batches in different levels to avoid layout thrashing, and to automatically process batches asynchronously to enable multiple synchronous calls being grouped into a pending batch.

    Being able to process a batch in levels is important when different types of operations, that are to be processed in a specific order (usually to avoid layout thrashing), needs to be grouped together in a batch.
    For example, a function that doubles an element's width and reads the new calculated height benefits by being batch processed in three levels: reading the width, mutating the width, and reading the height.
    The following is an example implementation of such function that uses the leveled batch processor:

    \begin{lstlisting}[gobble=6,label={},caption={},captionpos=b]]
      var batchProcessor = ...;

      function doubleWidth(element, callback) {
        // First level: reading the width.
        var width = element.offsetWidth;
        var newWidth = (width * 2) + "px";

        // Second level: mutating the width.
        // This is executed in level 0 of the batch.
        batchProcessor.add(0, function mutateWidth() {
          element.style.width = newWidth;
        });

        // Third level: reading the height.
        // This is executed in level 1 of the batch,
        // after level 0. Changing the level number 
        // from "1" to "0" results in layout thrashing.
        batchProcessor.add(1, function readHeight() {
          var height = element.offsetHeight;
          callback(height);
        });
      }
    \end{lstlisting}

    It should be noted that the first level is executed asynchronously by the function, and not handled by the actual batch processor.
    Since each batch is delayed to execute asynchronously, all synchronous calls of the method is grouped into a pending batch.
    If the batch would not automatically be delayed, layout thrashing would occur when the method is called multiple times.
    This results in a 45-fold speedup, when applied to 1000 elements, of the function compared to not processing the batch in levels.
    It also results in a simple API that allows multiple synchronous calls without causing layout thrashing, like so:

    \begin{lstlisting}[gobble=6,label={},caption={},captionpos=b]]
      var elements = [...];
      elements.forEach(function (element) {
        doubleWidth(element, function (height) {
          ...
        });
      });
    \end{lstlisting}

    The \code{activate} method of \elq{} is implemented similarly so it may also be called multiple times synchronously, without performance penalties, like the following example:

    \begin{lstlisting}[gobble=6,label={},caption={},captionpos=b]]
      var elements = [...];
      elements.forEach(elq.activate);
    \end{lstlisting}
  
  \subsection{Element Resize Detection}\label{sec:imp_erd}
    Unfortunately, there is no standardized resize event for arbitrary elements \cite{w3c_dom2_events}.
    A naive approach to detecting element resize events is to have a script continously check elements if they have resized given some interval (polling).
    This approach is appealing because it does not mutate the \gls{DOM}, supports arbitrary elements, and it provides excellent compatibility.
    However, in order to prevent the \gls{responsive} elements lagging behind the size changes of the user interface, polling needs to be performed quite frequently.
    The problem is that each poll forces the layout queue to be flushed since the computed style of elements needs to be retrieved in order to know if elements have resized or not \cite{elq-thesis}.
    Since the polling is performed all the time the overall page performance is decreased even if the page is idle, which is undesired especially for mobile devices running on battery.



    It is desired to instead have an event-based approach that only performs additional computations when an actual element resize has happened.
    This is achieved by the resize detection subsystem of \elq{} by using two independent injecting approaches, both originally presented by \cite{backalley}.
    These appraoches are limited to non-void elements, i.e., elements that may have content.
    It is a reasonable limitation since void elements can easily be wrapped with non-void elements without affecting the page visually.
    Like the batch processor, the resize detection subsystem is also implemented as a stand-alone project.\footnote{\url{https://github.com/wnr/element-resize-detector}}

    \paragraph{Object-based resize detection}
    Only documents emit resize events in modern browsers and therefore such events can only be observed for frame elements (since a frame \gls{element} has its own \gls{document}).
    This approach injects \code{object} elements into the target element, which can be listened to resize events since \code{object} elements are frames.
    The \code{object} is styled so that it always matches the size of the target \gls{element} and so that it does not affect the page visually.
    This approach has good browser compatability and excellent resize detection performance, but imposes severe performance impacts during injection since \code{object} elements use a significant amount of memory as shown in Section~\ref{sec:eval-perf}.

    \paragraph{Scroll-based resize detection}
    This approach injects an \gls{element} that contains multiple overflowing elements that listen to scroll events.
    The overflowing elements are styled so that \code{scroll} events are emitted when the target \gls{element} is resized.
    For detecting when the target \gls{element} shrinks, two elements are needed; one for handling the scrollbars and one for causing them to scroll.
    Similarly, for detecting when the target \gls{element} expands, two elements are needed in the same way.
    As this solution only injects \code{div} elements, it offers greater opportunities for optimizations.
    The main algorithm that is performed when an element $e$ is to be observed for resize events is the following:
    \begin{enumerate}
      \item\label{itm:erd-algo-original-scroll-1} Get the computed style of $e$.
      \item\label{itm:erd-algo-original-scroll-2} If the element is positioned (i.e., \code{position} is not \code{static}) the next step is \ref{itm:erd-algo-original-scroll-4}.
      \item\label{itm:erd-algo-original-scroll-3} Set the position of $e$ to be \code{relative}. Here additional checks can be performed to warn the developer about unwanted side effects of doing this.
      \item\label{itm:erd-algo-original-scroll-4} Create the four elements needed (two for detecting when $e$ shrinks, and two for detecting when $e$ expands) and attach event handlers for the \code{scroll} event of the elements.
                                                  When the elements have been styled and configured properly, they are added as children to an additional container element that is injected into $e$.
      \item\label{itm:erd-algo-original-scroll-5} The current size of $e$ is stored and the scrollbars of the injected elements are positioned correctly.
      \item\label{itm:erd-algo-original-scroll-6} The algorithm waits for the \code{scroll} event handlers to be called asynchronously by the \gls{layout engine} (they are called since the previous step repositioned the scrollbars).
                                                  When the handlers have been called, the injection is finished and observers can be notified on resize events of $e$ when \code{scroll} events occur.
    \end{enumerate}

    Layout thrashing can be avoided by using the leveled batch processor described in Section~\ref{sec:imp_batch_processor}, which results in a significant performance improvement as shown in Section~\ref{sec:eval-perf}.
    The algorithm steps are batch processed in the following levels:
    \begin{enumerate}
      \item\label{itm:erd-algo-scroll-level-1}
        \textbf{The read level:}
        Step \ref{itm:erd-algo-original-scroll-1} is performed to obtain all necessary information about $e$.
        The information is stored in a shared state so that all other steps can obtain the information without reading the \gls{DOM}.
      \item\label{itm:erd-algo-scroll-level-2}
        \textbf{The mutation level:}
        Steps \ref{itm:erd-algo-original-scroll-2}, \ref{itm:erd-algo-original-scroll-3} and \ref{itm:erd-algo-original-scroll-4} are performed, which mutate the \gls{DOM}.
        All mutations performed in this level can be queued by layout engines.
      \item\label{itm:erd-algo-scroll-level-3}
        \textbf{The forced layout level:}
        Step \ref{itm:erd-algo-original-scroll-5} is performed, which forces the some layout engines to perform a layout.
    \end{enumerate}

    Since repositioning a scrollbar in some layout engines forces a layout, such operations need to be performed after that all other queueable operations have been executed.
    Therefore, step~\ref{itm:erd-algo-original-scroll-5} is performed in level~\ref{itm:erd-algo-scroll-level-3} as the last step.
    Even though some layout engines are unable to queue the repositing of scrollbars, it is still beneficial to batch process the algorithm since only pure layouts need to be performed (instead of having to recompute styles and synchronize the \gls{DOM} and render trees before each layout).
    As step~\ref{itm:erd-algo-original-scroll-6} is performed by the \gls{layout engine} asynchronously and does not interact with the \gls{DOM}, it does not need to be batch processed.

\section{Empirical evaluation}\label{sec:eval}
  \subsection{Performance}\label{sec:eval-perf}
    Only the performance of the element resize detection system has been performed.
    This due to the fact that detecting element resize events entails the significant performance penalties of the library.
    Also, it is hard to compare performance results of related libraries since the functionality is different.
    Fortunately, element resize detection is the common denominator of all automatic libraries and the results of this system can be compared faithfully.
    Measurements and graphs show evaluations performed in Chrome version 42 unless stated otherwise.

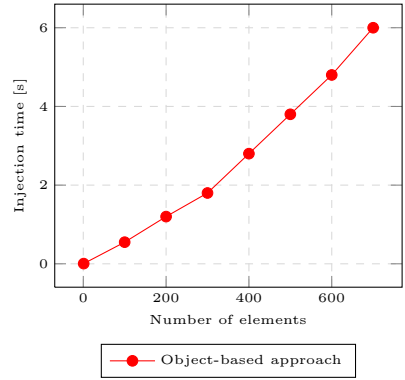
\begin{figure}[h]
      \tiny
      \begin{center}
        \begin{minipage}[t]{.35\textwidth}
          \vspace{0pt}
          \centering
            \begin{tikzpicture}
              \begin{axis}[
                  width=\textwidth, 
                  grid=major, 
                  grid style={dashed,gray!30}, 
                  xlabel=Number of elements, 
                  ylabel=Injection time,
                  y unit=s, 
                  legend style={at={(0.5,-0.20)},anchor=north} 
                ]
                \addplot+[red, mark options={red}] table[x=n elements,y=injection time,col sep=comma] {./data/erd-object-original.csv};
                \legend{Object-based approach}
              \end{axis}
            \end{tikzpicture}
        \end{minipage}%
      \caption{The injection performance of the object-based approach.}
      \label{fig:erd-original-object}
      \end{center}
    \end{figure}

    The object-based approach (as presented by \cite{backalley}) performs well when detecting resize events, which it does with a delay of 30 ms for 100 elements.
    However, the injection performance is not great as presented in figure~\ref{fig:erd-original-object}.
    As shown by the graph, the injection can be performed with adequate performance as long as the number of elements is low.
    The approach does not scale well as the number of elements increases.
    This is probably due to the fact that the heap memory usage grows roughly by 0.55 MB per element.

    \begin{figure*}[h]
      \tiny
      \begin{center}
        \begin{minipage}[t]{.5\textwidth}
          \vspace{0pt}
          \centering
            \begin{tikzpicture}
              \begin{axis}[
                  yticklabel style={
                          /pgf/number format/fixed,
                          /pgf/number format/precision=5
                  },
                  scaled y ticks=false,
                  width=\textwidth, 
                  grid=major, 
                  grid style={dashed,gray!30}, 
                  xlabel=Number of elements, 
                  ylabel=Injection time,
                  y unit=s, 
                  legend style={at={(0.5,-0.20)},anchor=north} 
                ]
                \addplot+[blue, mark=diamond*, mark options={blue}] table[x=n elements,y=injection time,col sep=comma] {./data/erd-scroll-elq.csv};
                \addlegendentry{ELQ scroll-based solution}
              \end{axis}
            \end{tikzpicture}
        \end{minipage}%
        \begin{minipage}[t]{.5\textwidth}
          \vspace{0pt}
          \centering

          \begin{tikzpicture}
            \begin{axis}[
                width=\textwidth, 
                grid=major, 
                grid style={dashed,gray!30}, 
                xlabel=Number of elements, 
                ylabel=Injection time,
                y unit=s, 
                legend style={at={(0.5,-0.20)},anchor=north} 
              ]
              \addplot+[red, mark options={red}] table[x=n elements,y=injection time,col sep=comma] {./data/erd-object-original.csv};
              \addplot+[orange, mark options={orange}] table[x=n elements,y=injection time,col sep=comma] {./data/erd-scroll-original.csv};
              \addplot+[blue, mark options={blue}] table[x=n elements,y=injection time,col sep=comma] {./data/erd-scroll-elq.csv};
              \addplot[dashed,red,domain=1:1500,samples=100] {5.567042796*10^(-6)*x^2 + 4.680174396*10^(-3)*x + 6.495754669*10^(-3)};
              \addplot[dashed,orange,domain=1:1500,samples=100] {4.533208873*10^(-6)*x^2 + 6.320377596*10^(-4)*x + 2.289566005*10^(-2)};
              \addplot[dashed,blue,domain=1:1500,samples=100] {4.071740702*10^(-8)*x^2 + 1.823749404*10^(-4)*x + 1.527042267*10^(-2)};
              \addlegendentry{Object-based solution}
              \addlegendentry{Scroll-based solution}
              \addlegendentry{ELQ scroll-based solution}
            \end{axis}
          \end{tikzpicture}
        \end{minipage}
      \caption{The left graph shows the injection time of the \elq{} scroll-based approach. The right graph shows all three approaches, including graph predictions by polynomial regression.}
      \label{fig:erd-elq-scroll}
      \end{center}
    \end{figure*}

    As the scroll-based approach (as presented by \cite{backalley}) does not inject \code{object} elements the memory footprint is reduced significantly, which improves the injection performance.
    The amount of used memory is too low for reliable measurements.
    See figure~\ref{fig:erd-elq-scroll} for graphs that show how the \elq{} scroll-based approach performs compared to the other two approaches.
    As evident in the figure, the optimized \elq{} approach has significantly reduced injection times.
    It achieves a 37-fold speedup compared to the object-based approach and a 17-fold speedup compared to the scroll-based approach when preparing 700 elements for resize detection.
    It also performs well when detecting resize events, which it does with a delay of 25 ms for 100 elements.

    \begin{table}[ht]\center
      \tiny
      \begin{tabular}[t]{ l l l l l l l }
        \multirow{2}{*}{Browsers} & \multicolumn{2}{c}{Injection} & \multicolumn{2}{c}{Resize detection} \\
        & scroll & object & scroll & object \\
        \hline
        Chrome v. 42                & 30 ms   & 550 ms    & 25 ms    & 20 ms  \\
        Firefox v. 40               & 150 ms  & 1000 ms   & 70 ms    & 30 ms  \\
        Safari v. 9                 & 100 ms  & 400 ms    & 30 ms    & 20 ms  \\
        Internet Explorer v. 11     & 350 ms  & 6700 ms   & 100 ms   & 80 ms  \\
        iOS Safari v. 9             & 350 ms  & 1600 ms   & 150 ms   & 60 ms  \\
        Android v. 5 Chrome v. 39   & 40 ms   & 1000 ms  & 20 ms     & 10 ms  \\
      \end{tabular}
      \caption{Performance of \elq{}'s two resize detection strategies, operating on 100 elements.}
      \label{table:erd-layout-engines}
    \end{table}

    \elq{} uses the object-based approach as a fallback for legacy browsers.
    Therefore the performance of the \elq{} resize detection system is at minimum as performant as related approaches.
    See table~\ref{table:erd-layout-engines} for the performance of \elq{}'s two resize detection strategies in different browsers.
    As shown in the table, the scroll-based approach is suitable for reduced page load, but may in some cases be preferred for better resize detection performance.

  \subsection{Case studies}
    In this section we aim to provide answers to the following questions:
    \begin{itemize}
      \item How can \elq{} be used to modularize existing responsive code bases?
      \item How much effort is this modularization?
    \end{itemize}

    In order to answer the questions, we have adapted the popular Bootstrap framework\footnote{This evaluation uses Bootstrap version 3.3.2.} to use element queries instead of media queries.
    According to its website, ``Bootstrap is the most popular HTML, CSS, and JS framework for developing responsive, mobile first projects on the web.''~\cite{bootstrap}
   
    In order to modularize Bootstrap, we redefine the behavior of its responsive elements so that they no longer respond to the viewport but to enclosing container elements.
    The following observation guides our modularization: all responsive elements should respond to their closest enclosing \code{container} or \code{container-fluid} element.
    Both classes are used in \gls{Bootstrap} to define new parts of a page (e.g., a grid is required to have a container ancestor).
    We also enable them to be nestable, which is important to satisfy the requirement of composable modules.
    The breakpoints of the container elements are defined using the \code{elq-breakpoints} API (see Section~\ref{sec:elq-api}).
    Since the Bootstrap API uses a predefined set of breakpoints, they are all added to the container elements with JavaScript.

    According to this design, we convert all responsive elements of Bootstrap to \code{elq-mirror} elements, since they need to mirror the breakpoints of the nearest ancestor \code{elq-breakpoints} element.
    Since container elements may be nested, they have both the \code{elq-breakpoints} and \code{elq-mirror} behavior.

    The breakpoints of Bootstrap are defined as the following constants:\footnote{The Bootstrap CSS is generated using the LESS preprocessor~\cite{lesscss} whose syntax we use.}
    \begin{lstlisting}[gobble=6,label={code:bootstrap-less-breakpoints},caption={},captionpos=b]]
      @screen-sm-min: 480px;
      @screen-md-min: 992px;
      @screen-lg-min: 1200px;
    \end{lstlisting}

    The following example shows how Bootstrap's style definitions are changed from using media queries to using \elq{}'s element queries:
    \begin{lstlisting}[gobble=6,label={code:bootstrap-less-breakpoints-usage},caption={},captionpos=b]]
      /* File "less/grid.less" of Bootstrap. */

      // Original Bootstrap using media queries.
      .container {
        @media (min-width: @screen-sm-min) {
          width: @container-sm;
        }
        ...
      }

      // ELQ Bootstrap using element queries.
      .container {
        &.elq-min-width-@{screen-sm-min} {
          width: @container-sm;
        }
        ...
      }
    \end{lstlisting}

    By using the power of preprocessors, \gls{ELQ} element queries become as pleasant to work with as \gls{media queries}.
    In fact, only about 0.6\% of the style code (LESS syntax) need to be altered.
    Most changes are similar to the one shown above, which replaces the media query syntax with the \gls{ELQ} element queries syntax.
    This is especially advantageous when keeping a forked project up to date with the original project, as fewer diverged lines implies a lowered risk of merge conflicts.


    In summary we have shown that it is easy to adapt existing responsive code to use \elq{}'s element queries instead of media queries.
    With only a small number of changes, the widely used Bootstrap framework can be modularized.

    \paragraph{Industrial use of \elq{}}
    In addition to the Bootstrap case study, we have been gathering experience with the application of \elq{} in large financial applications at EVRY.
    Our practical experience shows that complex applications require a variety of features to be supported by element queries.
    Such features can be provided effectively by \elq{} plugins.

\begin{table*}[ht!]\center
    \tiny
    \begin{tabular}[t]{ p{3cm} l l l l l }
      Implementation & Syntax & Resize detection & Page dynamism  & Composability & Cycle detection \\
      \hline
       MagicHTML \cite{eq_imp_magichtml}                          &                       Custom CSS  &   - &                             Static &     -                  & -  \\
       EQCSS \cite{eq_imp_eqcss}                                  &                       Custom CSS  &   Viewport only &                 Dynamic &    Full support       & -  \\
       Element Media Queries \cite{eq_imp_prollyfill-min-width}   &                       Custom CSS  &   Non-void elements &             Dynamic &    -                  & -  \\
       Localised CSS \cite{eq_imp_localised-css}                  &                       Custom CSS  &   Arbitrary elements &            Dynamic &    -                  & -  \\
       Grid Style Sheets 2.0 \cite{eq_imp_gss}                    &                       Custom CSS  &   Arbitrary elements &            Dynamic &    Partial support    & -  \\
       Class Query \cite{eq_imp_classquery}                       &                                 - &   - &                             Static &     -                  & -  \\
       breakpoints.js \cite{eq_imp_breakpointsjs}                 &                                 - &   Viewport only &                 Dynamic &    -                  & -  \\
       MediaClass \cite{eq_imp_mediaclass}                        &                                 - &   Viewport only &                 Dynamic &    -                  & -  \\
       ElementQuery \cite{eq_imp_elementquery}                    &                                 - &   Viewport only &                 Dynamic &    -                  & -  \\
       Responsive Elements \cite{eq_imp_responsive-elements}      &                                 - &   Viewport only &                 Dynamic &    -                  & -  \\
       SickleS \cite{eq_imp_sickles}                              &                                 - &   Viewport only &                 Dynamic &    -                  & -  \\
       Responsive Elements \cite{eq_imp_responsive-elements-2}    &                                 - &   Viewport only &                 Dynamic &    -                  & -  \\ 
       breaks2000 \cite{eq_imp_breaks2000}                        &                                 - &   Viewport only &                 Dynamic &    -                  & -  \\
       eq.js \cite{eq_imp_eqjs}                                   &                                 - &   Viewport only &                 Dynamic &    -                  & -  \\
       Element Queries \cite{eq_imp_element-queries}              &                                 - &   Non-void elements &             Dynamic &    -                  & -  \\
       CSS Element Queries \cite{eq_imp_css-element-queries}      &                                 - &   Non-void elements &             Dynamic &    -                  & -  \\
       Selector queries and responsive containers \cite{eq_imp_selector_queries}                  & - &   Arbitrary elements &            Dynamic &    -                  & -  \\
       \elq{}                                                                                     & - &   Non-void elements &             Dynamic &    Full support       & Yes \\
    \end{tabular}
    \caption{Classification of related approaches to modular RWD.}
    \label{table:approaches-classifications}
  \end{table*}

\section{Discussion}\label{sec:discussion}

  \subsection{Limitations}
    Inherent to all current implementations of element queries is that the conditional style is applied ``one layout behind''.
    Since a layout pass needs to have been performed in order for an element to change size, the conditional styles defined by the element queries cannot be applied until next layout.
    Therefore, the element will display invalid design until another layout has been performed.
    The flash of invalid design is usually so short that users do not notice it, but in some cases developers need to work around this issue to avoid more apparent results.

    Another caveat is presented by the element resize detection approaches, as they mutate the DOM.
    Developers need to be aware of this as CSS selectors and JavaScript may also match the injected elements.
    This is easily avoided by good practices.

    It should be noted that all limitations described only affects the elements that uses the element queries functionality.
    \elq{} does not impose potential problems to other parts of the DOM other than where applied explicitly.

    Currently \elq{} only supports breakpoints for the width and height element properties, as it has been identified as the general use case \cite{elq-thesis}.
    In the future, we aim to support plugins to define custom breakpoint properties.

  \subsection{Standardization}
    It is stated on the W3C's www-style mailing list \cite{w3c_eq_mail} by Zbarsky of Mozilla, Atkins of Google and Sprehn of Google that element queries are infeasible to implement without restricting them.
    By limiting element queries to specially separated container elements that can only be queried by child elements, many of the problems are resolved \cite{ricg_irc_log,ricg_issue_viewport}.
    Therefore, the Responsive Issues Community Group (RICG) is currently investigating the possibility of standardizing \emph{container} queries.

    Unfortunately, even such limited container queries requires significant effort to implement due to the complex changes to browsers required \cite{ricg_issue_viewport}.
    Atkins argues that a full implementation that avoids the double layout issue is unlikely to be implemented, and therefore it might be wiser to pursue sub-standards that aids third-party solutions instead.
   
    In the future, we hope that \elq{} may use the aiding sub-standards pushed by RICG, to achieve greater flexibility and performance.
    A standardized resize event would enable us to avoid injecting elements, and to reduce the code base of \elq{} significantly.
    Support for custom at-rules/selectors would also enable us to define a more natural API in CSS.
    Finally, being able to tell elements to ignore children while computing their size would decrease the need for cycle detection.

\section{Related Work}\label{sec:related}
  Table~\ref{table:approaches-classifications} attempts to classify all existing approaches, of modular RWD, known to us. 
  We discuss these approaches according to two different aspects: (a) syntax extensions and (b) resize detection.

  \paragraph{Syntax extensions}
  The libraries \cite{eq_imp_magichtml,eq_imp_eqcss,eq_imp_prollyfill-min-width,eq_imp_localised-css,eq_imp_gss} have in common that they require developers to write custom \gls{CSS}, unlike \elq{}.
  Since they do not conform to the \gls{CSS} standard, new features are supported through custom \gls{CSS} parsed using JavaScript.
  As shown by \cite{eq_imp_eqcss,eq_imp_gss} quite advanced features can be implemented this way.
  Additionally, adding new \gls{CSS} features implies that it is possible to implement a solution to element queries that does not require any changes to the \gls{HTML}, which may be preferable since all styling then can be written in \gls{CSS}.
  However, there are numerous drawbacks with libraries that require custom \gls{CSS}.
  Extending the CSS syntax violates the requirement of compatability and also introduces a compilation step which decreases the performance \cite{elq-thesis}.

  \paragraph{Resize detection}
  The libraries \cite{eq_imp_eqcss,eq_imp_breakpointsjs,eq_imp_mediaclass,eq_imp_elementquery,eq_imp_responsive-elements,eq_imp_sickles,eq_imp_responsive-elements-2,eq_imp_breaks2000,eq_imp_eqjs} simply observe the \gls{viewport} resize event, which may be enough for static pages, but not enough to satisfy the requirements of reusable responsive modules \cite{elq-thesis}.
  Approach \cite{eq_imp_classquery} does not detect resize events at all.
  Like \elq{}, \cite{eq_imp_localised-css,eq_imp_selector_queries,eq_imp_prollyfill-min-width,eq_imp_gss,eq_imp_element-queries,eq_imp_css-element-queries} observe \emph{elements} for resize events.
  The libraries \cite{eq_imp_localised-css,eq_imp_selector_queries} use polling while \elq{} and \cite{eq_imp_prollyfill-min-width,eq_imp_gss,eq_imp_element-queries,eq_imp_css-element-queries} use different injection approaches, as described in Section~\ref{sec:imp_erd}.
  As shown in Section~\ref{sec:eval-perf}, the injection approaches used by related libraries have significanly less performance than the element resizing detection system used in \gls{ELQ}.

  \paragraph{Constraint-based CSS}
  CCSS~\cite{badros1999constraint} proposes a more general and flexible alternative to CSS.
  As the name suggests, the idea of \gls{CCSS} is to layout documents based on constraints.
  According to its authors, the constraint-based approach provides extended features and reduced complexity compared to CSS.
  To solve the constraints CCSS uses the Cassowary constraint solving algorithm~\cite{BadrosBS01}.

  The Grid Style Sheets library \cite{eq_imp_gss} builds upon the ideas of \gls{CCSS} and uses a \gls{JavaScript} port~\cite{cassowary_js} of Cassowary to solve the constraints at runtime.
  While not directly offering element queries, the library enables the possibility to conditionally style elements by \gls{element} criteria and thus makes it a good candidate to solve the problem of \gls{responsive} modules.
  However, the library has two major issues: performance and browser compatibility \cite{gss_issue}.
  One approach to resolve both issues is to precompute the layout in a compilation step at the server.
  However, precompiling styles implies static layouts.
  The authors discuss other approaches~\cite{gss_issue} that would increase the performance while limiting the dynamism of page layout.
  In contrast, \elq{} only considers element queries, but without these limitations and with higher performance.

\section{Conclusion}\label{sec:conclusion}

  Responsive Web Design (RWD) enables web applications to adapt to the characteristics of different devices, which is achieved using CSS media queries.
  However, using media queries it is impossible to create responsive applications in a modular way, because responsive elements then always depend on the global context.

  This paper extends RWD to also include responsive modules through element queries.
  We present \elq{}, an open-source implementation of our approach, that conforms to the current standards of HTML, CSS and JavaScript.
  It enables developers to create responsive modules that are independent of their context, and a way to encapsulate their conditional style rules.
  The element resize detection of \elq{}, used to automatically evaluate element queries on changes of responsive elements, performs up to 37x better than previous algorithms.

  Using a case study based on the popular Bootstrap framework we show that large code bases using media queries can be converted to using \elq{}'s element queries with little effort.
  Changing only about 0.6\% of the LOC of style related code was sufficient to enable the use of Bootstrap in responsive modules.
  We also report on first commercial usage of \elq{}.
  
  We believe \elq{} is an important contribution to realizing a modular form of element queries, in particular since standardization bodies like the RICG do not intend to standardize a complete solution.
  In the future we intend to improve \elq{} by using forthcoming standards developed by the RICG to avoid some current limitations.


\newpage
\section{Acknowledgments}
The authors would like to thank EVRY for sponsoring the \elq{} project including the supporting projects for element resize detection and batch processing.

%
\bibliographystyle{abbrv}
\bibliography{elq}  
%
%
\end{document}